# Gravity Field and Electromagnetic Field

## Finite Geometrical Field Theory of Matter Motion Part Two

Xiao Jianhua

Natural Science Foundation Research Group, Shanghai Jiaotong University

Shanghai, P.R.C

**Abstract:** Gravity field theory and electromagnetic field theory are well established and confirmed by experiments. The Schwarzschild metric and Kerr Metric of Einstein field equation shows that the spatial differential of time gauge is the gravity field. For pure time displacement field, when its spatial differentials are commutative, conservative fields can be established. When its spatial differentials are non-commutative, Maxwell electromagnetic field equations can be established. When the contra-covariant is required for the non-commutative field, both Lorentz gauge and Coulomb gauge are derived in this research. The paper shows that the light is a special matter in that the addition of its Newtonian mass and its Coulomb electric charge is zero. In fact, this conclusion is true for the electromagnetic wave in vacuum. For the conservative field, the research shows that once the mass density and the Coulom charge dendity are given, the macro spacetime feature is completely determined. Both of them are intrisinc features of macro matter in cosimic background. However, for the cosmic ages old events, the spatial curvature may be cannot be ignored. On this sense, the oldest gravity field has the largest curvature of space. This point is very intrinsic for astronomy matters.

### 1. Introduction

In the paper "Inertial System and Special Relativity"[1], the symmetric matter motion transformation is discussed.

For Newtonian matter motion, the vacuum is taken as a special matter with one intrinsic physical parameter-light speed $c$. For any matter, it can be identified with co-moving coordinators with local varying geometry. The local varying geometry represents the motion of matter in consideration. The motion equation of matter is determined by cosmic environment where the matter existing, where two intrinsic physical parameters are introduced to define matter as $(\lambda, \mu)$.

When the time gradient of time displacement is forced to be zero, the Newton's mechanical equation is gotten. It shows that the Newton's mass is determined by the shear feature of space-time continuum. The Newton's mass is expressed as $\rho = \mu/c^2$.

When the matter has no macro spatial motion, such as gravity field or electronic field, the time displacement field is introduced to describe the matter motion. One finds that gravity field, electro-magnetic field, and quantum field are time gradient field. They are related with Newton's mechanics in intrinsic sense. When there is no electrical charge quantity, the gravity mass and the inertial mass in Newton's mechanics are the same.

The matter has three typical existing forms: traveling wave, localized harmonic vibrating particle, and exponential expanding or decaying matter field.

The commutable matter motion defines conservative field. The non-commutable matter motion defines quantum field. The wave particle duality is explained by deformable matter motion.

In this part, pure time displacement field will be discussed. In this case the motion transformation is non-symmetric. The research shows that the electromagnetic field can be defined by the pure time displacement field, the Maxwell equations are derived in this part.

Firstly, the basic equations are reviewed simply.

The matter motion in four-dimensional spacetime continuum is defined by:

$$\begin{vmatrix} \vec{g}_1 \\ \vec{g}_2 \\ \vec{g}_3 \\ \vec{g}_4 \end{vmatrix} = \begin{vmatrix} 1+u^1|_1 & u^2|_1 & u^3|_1 & u^4|_1 \\ u^1|_2 & 1+u^2|_2 & u^3|_2 & u^4|_2 \\ u^1|_3 & u^2|_3 & 1+u^3|_3 & u^4|_3 \\ u^1|_4 & u^2|_4 & u^3|_4 & 1+u^4|_4 \end{vmatrix} \cdot \begin{vmatrix} \vec{g}_1^0 \\ \vec{g}_2^0 \\ \vec{g}_3^0 \\ \vec{g}_4^0 \end{vmatrix} \quad (1)$$

That is the initial basic vectors and current basic vectors meet transformation equation:

$$\vec{g}_i = F_i^j \vec{g}_j^0 \quad (2)$$

The matter motion transformation tensor $F_j^i$ is determined by equation:

$$F_i^j = u^j|_i + \delta_i^j \quad (3)$$

where, $|_i$ represents covariant derivative with respect to coordinator $x^i$, $\delta_i^j$ is an unit tensor.

The displacement field $u^i$ is also defined in the initial four-dimensional co-moving coordinator system. Therefore, the matter motion is expressed by displacement field $u^i$ measured in standard physical measuring system. Its three dimenssional case has been studied by Chen Zhida [2-3].

Generally, the matter motion transformation tensor is non-commutative. Its covariant index and contra-variant index are defined respects with to the initial four-dimensional co-moving coordinator system.

The action for matter motion is defined as:

$$Action = \int W(F_j^i) dx^1 dx^2 dx^3 dx^4 \quad (4)$$

where, $W$ is a general function.

The least action principle [4-8] gives the field motion equations as:

$$\sigma_j^i\big|_j = f^i \quad (5)$$

where,

$$\sigma_j^i = C_{jl}^{ik}(F_k^l - \widetilde{F}_k^l) \quad (6)$$

$$C_{jl}^{ik} = \frac{\partial^2 W}{\partial(\widetilde{F}_j^i)\partial(\widetilde{F}_l^k)} \quad (7)$$

$$\left.\frac{\partial W}{\partial (\tilde{F}_j^i)}\right|_j = -f^i \quad (8)$$

Usually, one does not care how the original matter motion state is in the absolute spacetime. One just takes the matter under discussion as a self-exist object. In this sense, the initial matter can be generally defined by select the initial co-moving coordinator system to make:

$$\tilde{F}_j^i = \delta_j^i \quad (9)$$

On this sense, the reference matter is taken to define the spacetime continuum where the matter under discussion is moving.

The principle of physical laws covariant invariance requires the general reference matter motion be coordinator independent. To meet these requirements, the tensor $C_{jl}^{ik}$ must be isotropic tensor. Therefore, one has:

$$C_{jl}^{ik} = \lambda \delta_j^i \delta_l^k + \mu \delta_l^i \delta_j^k \quad (10)$$

where, $\lambda$ and $\mu$ are the intrinsic feature of matter referring to reference matter. For Newton mechanics, the reference matter is vacuum or ether. For the isotropic tensor in mixture form please refer [9]. In standard physical theory, the mixed form is refused. However, the research shows that the mixture form is necessary to show the non-commutative feature of matter motion transformation tensor. This topic should treat in future rather at present.

For simplest Newton motion, the related physical form of deformation tensor and stress tensor are:

$$F_j^i - \delta_j^i = c^{-1} \cdot \begin{vmatrix} 0 & 0 & 0 & V^1 \\ 0 & 0 & 0 & V^2 \\ 0 & 0 & 0 & V^3 \\ V^1 & V^2 & V^3 & 0 \end{vmatrix} \quad (11)$$

$$\sigma_j^i = \mu_0 \cdot c^{-1} \cdot \begin{vmatrix} 0 & 0 & 0 & V^1 \\ 0 & 0 & 0 & V^2 \\ 0 & 0 & 0 & V^3 \\ V^1 & V^2 & V^3 & 0 \end{vmatrix} \quad (12)$$

The inertial mass density is defined as:

$$\rho = \mu_0 \cdot c^{-2} \quad (13)$$

It interprets the $\mu_0$ as the intrinsic inertial energy of matter.

The symmetric motion transformation in the form:

$$F_j^i - \delta_j^i = \begin{vmatrix} S_1^1 & 0 & 0 & \dfrac{\partial u^4}{\partial x^1} \\ 0 & S_2^2 & 0 & \dfrac{\partial u^4}{\partial x^2} \\ 0 & 0 & S_3^3 & \dfrac{\partial u^4}{\partial x^3} \\ \dfrac{V^1}{c} & \dfrac{V^2}{c} & \dfrac{V^3}{c} & \dfrac{\partial u^4}{c\partial t} \end{vmatrix} \qquad (14)$$

Where, $S_j^i$ are classical principle spatial strain components. The symmetry condition requires that:

$$\frac{\partial u^1}{c\partial t} = \frac{\partial u^4}{\partial x^1}, \quad \frac{\partial u^2}{c\partial t} = \frac{\partial u^4}{\partial x^2}, \quad \frac{\partial u^3}{c\partial t} = \frac{\partial u^4}{\partial x^3} \qquad (15)$$

For simplicity, the $u^4$ at here is viewed as physical component (in fact it is not).

The corresponding physical stress components are:

$$\sigma_j^i = (\lambda + \mu) \cdot (S_1^1 + S_2^2 + S_3^3 + \frac{\partial u^4}{c\partial t})\delta_j^i$$

$$+ \mu \cdot \begin{vmatrix} 0 & 0 & 0 & \dfrac{\partial u^4}{\partial x^1} \\ 0 & 0 & 0 & \dfrac{\partial u^4}{\partial x^2} \\ 0 & 0 & 0 & \dfrac{\partial u^4}{\partial x^3} \\ V^1/c & V^2/c & V^3/c & 0 \end{vmatrix} \qquad (16)$$

For non-compressible matter defined by: $\Delta = (S_1^1 + S_2^2 + S_3^3) = 0$, the basic matter motion forms are: matter wave, wave-particle quantum, and conservative field (gravitation field or static electrical field).

Although there are many points should be cleared, to express the main ideas of this research, the paper will go straight a way. Only when the whole research is carefully polished at later time, some ambiguity can be effectively removed.

## 2. Matter Motion Without Macro Spatial Motion

There exists such a kind of matter motion that time displacement gradient are non-zero while the spatial displacement gradient are all zeros. *Note that the physical components of time $dt$ will be used in physical motion equation in the sense that: $dx^4 = cdt$. Such a treatment will make the physical meaning of $u^4$ be more clear (the cost is only the mathematical form). The covariant derivative of stress for $\dfrac{\partial}{\partial x^4}$ is simply replaced by $\dfrac{\partial}{cdt}$. This is for simplicity sake, as the purpose of the paper is to show the physical relation between various physical fields.* For the matter motion without macro spatial motion, the deformation of space-time continuum is:

$$F^i_j - \delta^i_j = \begin{vmatrix} 0 & 0 & 0 & \dfrac{c\partial u^4}{\partial x^1} \\ 0 & 0 & 0 & \dfrac{c\partial u^4}{\partial x^2} \\ 0 & 0 & 0 & \dfrac{c\partial u^4}{\partial x^3} \\ 0 & 0 & 0 & \dfrac{\partial u^4}{\partial t} \end{vmatrix} \quad (17)$$

The stress tensor is:

$$\sigma^i_j = (\lambda + \mu) \cdot \dfrac{\partial u^4}{\partial t} \delta^i_j + \mu \cdot c \cdot \begin{vmatrix} 0 & 0 & 0 & \dfrac{\partial u^4}{\partial x^1} \\ 0 & 0 & 0 & \dfrac{\partial u^4}{\partial x^2} \\ 0 & 0 & 0 & \dfrac{\partial u^4}{\partial x^3} \\ 0 & 0 & 0 & 0 \end{vmatrix} \quad (18)$$

Its stress field is composed of an isotropic stress tensor and a non-symmetrical stress tensor. The isotropic stress tensor shows the matter may be a pure continuum matter field (be named as field matter, here after). This is significantly different from Newtonian matter in that the Newtonian matter is related with symmetrical stress tensor.

The equation (5) shows that field matter motion equations are:

$$\begin{aligned} f^1 &= (\lambda + \mu) \dfrac{\partial^2 u^4}{\partial t \partial x^1} + \mu \dfrac{\partial^2 u^4}{\partial x^1 \partial t} \\ f^2 &= (\lambda + \mu) \dfrac{\partial^2 u^4}{\partial t \partial x^2} + \mu \dfrac{\partial^2 u^4}{\partial x^2 \partial t} \\ f^3 &= (\lambda + \mu) \dfrac{\partial^2 u^4}{\partial t \partial x^3} + \mu \dfrac{\partial^2 u^4}{\partial x^3 \partial t} \\ f^4 &= \dfrac{\lambda + \mu}{c} \dfrac{\partial^2 u^4}{(\partial t)^2} \end{aligned} \quad (19)$$

The commutative and non-commutative cases will be discussed bellow.

## 3. Conservative Field Matter Motion Equation (Commutative Case)

If the partial differential of time displacement is commutable about space and time differentiation, the matter will behave as conservative field matter. As the commutability of space and time differentiation will make the equation (19) be rewritten as:

$$\begin{aligned} f^1 &= (\lambda + 2\mu) \dfrac{\partial}{\partial x^1}(\dfrac{\partial u^4}{\partial t}) = \tilde{\rho} c^2 \dfrac{\partial A}{\partial x^1} \\ f^2 &= (\lambda + 2\mu) \dfrac{\partial}{\partial x^2}(\dfrac{\partial u^4}{\partial t}) = \tilde{\rho} c^2 \dfrac{\partial A}{\partial x^2} \\ f^3 &= (\lambda + 2\mu) \dfrac{\partial}{\partial x^3}(\dfrac{\partial u^4}{\partial t}) = \tilde{\rho} c^2 \dfrac{\partial A}{\partial x^3} \\ f^4 &= \dfrac{\lambda + \mu}{c} \dfrac{\partial^2 u^4}{(\partial t)^2} = \dfrac{\lambda + \mu}{c} \dfrac{\partial A}{\partial t} = \tilde{q} c \dfrac{\partial A}{\partial t} \end{aligned} \quad (20)$$

where:

$$\tilde{\rho} = \dfrac{\lambda + 2\mu}{c^2} = \tilde{q} + \rho \,, \quad \tilde{q} = \dfrac{\lambda + \mu}{c^2} \quad (21)$$

$$\frac{\partial u^4}{\partial t} = A \tag{22}$$

And $\tilde{\rho}$ is defined as matter complete charge quality; scalar $A$ is defined as the potential of matter field. It can be seen that $A$ may represent gravity field and/or static electronic field. Based on this understanding, gravity field and electronic field are produced by time displacement gradient. For gravity field, the Schwarzschild's solution of gravity field [6-8] introduces Newton's gravity field with the help of time gauge differentiation.

The equation (20) has three typical solutions, which depend on the feature of cosmic force $f^4$ in the case that it is a constant. For the exact physical meaning of it will be discussed later (As it depends on the action function time variation, so it can be understood as energy at present. In the paper, it is roughly named as cosmic force without detailed discussion).

*(a). Absolute Time Velocity Solution: Gravity Field and Static Electric Field*

If $f^4 = 0$, that is there is no time acceleration, the solution is:

$$u^4 = At + B \tag{23}$$

Where $A$ is a spatial function. In this case, the spatial forces $f^i$ for $i = 1,2,3$ are:

$$f^i = \tilde{\rho} c^2 \frac{\partial^2 u^4}{\partial t \partial x^i} = \tilde{\rho} c^2 \frac{\partial A}{\partial x^i} \tag{24}$$

This is the conservative field in traditional mechanics, where, the potential of matter field $A$ is a spatial function. This includs electronic field and gravity field. Details about it please see the paper "Inertial System and Special Relativity"[1].

As $\tilde{\rho} = \frac{\lambda + 2\mu}{c^2} = \frac{\lambda + \mu}{c^2} + \frac{\mu}{c^2}$, according to our previous research in the part one of this full research paper, the $\frac{\mu}{c^2}$ is Newtonian inertial mass density. Therefore, the another part should be defined as the electric charge (Coulomb charge). The electric charge density may defined as $q = \lambda + \mu$, or $\tilde{q} = \frac{\lambda + \mu}{c^2}$. Under above definition, the inertial mass and the static electric charge are related in intrinsic sense as:

$$\tilde{q} - \rho = \frac{\lambda}{c^2}, \text{ or } q - \rho c^2 = \lambda \tag{25}$$

Now, it is clear that onec the mass density and the Coulom charge dendity are given, the macro matter feature is completely determined. There for both of them are intrisinc features of macro matter in cosimic background.

For massless inertial matter, $\mu = 0$, the third kind of matter conservative field can exist. That is the pure $\pi = \lambda/c^2$ charge field. Its spatial for field is:

$$f_\lambda^i = \lambda c^2 \frac{\partial^2 u^4}{\partial t \partial x^i} = \lambda c^2 \frac{\partial A}{\partial x^i} \tag{26}$$

As for electric charge-less matter will require the condition: $q = 0$, that means: $\lambda = -\mu$. In this sense, it is the negative matter, that is:

$$f_\lambda^i = -\mu \cdot c^2 \frac{\partial^2 u^4}{\partial t \partial x^i} = -\rho \cdot c^2 \frac{\partial A}{\partial x^i} \qquad (27)$$

The negative matter has been discussed in physics for long time. Here, the research views the negative matter the matter defined by $\lambda$ parameter. If one takes the matter and negative matter as the basic parameter, the electrical charge will be a derivative parameter (lost its basic position). As the physics has taken the mass and electrical charge as basic parameter, there is no need to introduce the negative matter concept (although it may be helpful for some cases).

*(b). Constant Time Acceleration Solution: Time Dependent Spatial Force*

If $f^4 > 0, q > 0$, or $f^4 < 0, q < 0$, one will get the time displacement:

$$u^4 = C \cdot \exp[\sqrt{\frac{f^4}{\tilde{q}c}} \cdot t] + D \cdot \exp[-\sqrt{\frac{f^4}{\tilde{q}c}} \cdot t] \qquad (28)$$

The spatial forces $f^i$ for $i = 1,2,3$ are:

$$f^i = \tilde{\rho} c^2 \frac{\partial^2 u^4}{\partial t \partial x^i} = \tilde{\rho} c^2 \sqrt{\frac{f^4}{\tilde{q}c}} \cdot \{\frac{\partial C}{\partial x^i} \cdot \exp[\sqrt{\frac{f^4}{\tilde{q}c}} \cdot t] - \frac{\partial D}{\partial x^i} \cdot \exp[-\sqrt{\frac{f^4}{\tilde{q}c}} \cdot t]\} \qquad (29)$$

This solution corresponds to basic particle field in quantum mechanics, where $C$ and $D$ are spatial function.

The spatial force is time increasing or decreasing. In essential sense, there is no steady solution. It confirms that basic particles can be produced or decayed.

*(c). Constant Time Deceleration Solution: Quantum Wave Spatial Force*

If $f^4 < 0, q > 0$, or $f^4 > 0, q < 0$, one will get the time displacement:

$$u^4 = E \cdot \exp[\tilde{j}\sqrt{\frac{f^4}{\tilde{q}c}} \cdot t] + F \cdot \exp[-\tilde{j}\sqrt{\frac{f^4}{\tilde{q}c}} \cdot t] \qquad (30)$$

Where, the $\tilde{j}$ in the right side of equation is the sign of imaginary number; $E$ and $F$ are spatial function.

The spatial forces $f^i$ for $i = 1,2,3$ are:

$$f^i = \tilde{\rho} c^2 \frac{\partial^2 u^4}{\partial t \partial x^i} = \tilde{j} \tilde{\rho} c^2 \sqrt{\frac{f^4}{\tilde{q}c}} \{\frac{\partial E}{\partial x^i} \cdot \exp[\tilde{j}\sqrt{\frac{f^4}{\tilde{q}c}} \cdot t] - \frac{\partial F}{\partial x^i} \cdot \exp[-\tilde{j}\sqrt{\frac{f^4}{\tilde{q}c}} \cdot t]\} \qquad (31)$$

This solution corresponds to self-rotation of particle or vibrating field. It shows how the electronic field and gravity field are coupled together in wave form.

In fact, for positive energy, the solution shows the motion of particles with negative electric charge.

The (b) and (c) cases show that the time feature of commutative field matter motion is determined by the electric charge and cosmic force. In fact, they are the basic form of molecular or atom quantum field, where the elctric charge play an important role. The frequency of quantum can be defined as:

$$v = \sqrt{\frac{f^4}{\tilde{q}c}} \tag{32}$$

When the cosmic force is physical component form, that is: $\tilde{f}^4 = cf^4$, the frequency is:

$$v = \sqrt{\frac{\tilde{f}^4}{\tilde{q}}} \tag{33}$$

As the basic unit of electric charge is a limit quantity $e = -1.6021892 \times 10^{-19} C$, the frequency can be expressed as:

$$v = \sqrt{\frac{1}{-e}} \cdot \sqrt{\frac{\tilde{f}^4}{N}} \tag{34}$$

where, $N$ is an integer number with the same sign with $\tilde{f}^4$. It means that for fixed $\tilde{f}^4$, the frequency is discrete rather than continous. This is caused by the discretity of electric charge.

On the other hand, it can be infereed that for the matter with very large electric charge quantity the frequency will be continuous. Hence, the quantum phenomenor is related with samll charge matter. Note that it is conservative quantum field.

## 4. Non-commutable Field Matter Motion Equation (Quantum Mechanics)

If the partial differential of time displacement is non-commutative about space and time differentiation, the matter will behave as quantum electromagnetic field matter. The equation (19) be rewritten as:

$$\begin{aligned} f^1 &= (\lambda + \mu) \frac{\partial}{\partial x^1} (\frac{\partial u^4}{\partial t}) = (\lambda + \mu) \frac{\partial A}{\partial x^1} + \mu \frac{\partial B_1}{\partial t} \\ f^2 &= (\lambda + \mu) \frac{\partial}{\partial x^2} (\frac{\partial u^4}{\partial t}) = (\lambda + \mu) \frac{\partial A}{\partial x^2} + \mu \frac{\partial B_2}{\partial t} \\ f^3 &= (\lambda + \mu) \frac{\partial}{\partial x^3} (\frac{\partial u^4}{\partial t}) = (\lambda + \mu) \frac{\partial A}{\partial x^3} + \mu \frac{\partial B_3}{\partial t} \\ f^4 &= \frac{\lambda + \mu}{c} \frac{\partial^2 u^4}{(\partial t)^2} = \frac{\lambda + \mu}{c} \frac{\partial A}{\partial t} \end{aligned} \tag{35}$$

where,

$$B_i = \frac{\partial u^4}{\partial x^i}, \quad i = 1,2,3 \tag{36}$$

As $A$ is defines the conservative matter field, the $B_i$ should be defined as the vector potential of matter field, such as magnetic field. It should be noted that the force produced by the vector potential of matter field is related with Newton's matter mass.

The equation (35) has three typical solutions, which depend on the feature of cosmic force $f^4$ in the case that it is a constant.

*(a). Maxwell Electromagnetic Field Equation (and Coherent Light)*

If the spatial force is zero, that is $f^i = 0$, one will get:

$$(\lambda + \mu)\frac{\partial A}{\partial x^1} + \mu\frac{\partial B_1}{\partial t} = 0$$

$$(\lambda + \mu)\frac{\partial A}{\partial x^2} + \mu\frac{\partial B_2}{\partial t} = 0 \quad (37)$$

$$(\lambda + \mu)\frac{\partial A}{\partial x^3} + \mu\frac{\partial B_3}{\partial t} = 0$$

One recognizes that the equation (38) is Maxwell's equation for electro-magnetic field in vacuum. This equation is a basic equation for quantum mechanics.

In vector form, the related equations are:

$$\vec{D} = (\lambda + \mu)\nabla A = \tilde{q}c^2 \nabla \frac{\partial u^4}{\partial t} \quad (38)$$

$$\vec{H} = \mu \nabla \times \vec{B} = \rho c^2 \nabla \times \nabla u^4, \quad \nabla \cdot \vec{H} = 0 \quad (39)$$

They are definition equations for electric field $\vec{E}$ and magnetic field $\vec{H}$.

In this sense, the equation (37) gives out:

$$\nabla \times \vec{D} + \frac{\partial \vec{H}}{\partial t} = 0 \quad (40)$$

From the fourth equation of (35), a $u^4$ solution function is available as discussed in previous sub-section *(Conservative Field Matter Motion Equation)*. For the simplest solutions in form (23), one has:

$$\nabla \cdot \vec{D} = \tilde{q}c^2 \nabla^2 \frac{\partial u^4}{\partial t} = \tilde{\sigma} \quad (41)$$

$$\nabla \times \vec{H} = \rho c^2 \nabla \times \nabla \times \nabla u^4 = \frac{\partial \vec{D}}{\partial t} + \vec{J} \quad (42)$$

where:

$$\frac{\partial \vec{D}}{\partial t} + \vec{J} = \rho c^2 [\nabla(\nabla^2 u^4) - \nabla^2(\nabla u^4)] = \rho c^2 [\nabla(\nabla \cdot \vec{B}) - \nabla^2 \vec{B}] \quad (43)$$

To make the problem simple, Coulomb gauge:

$$\nabla^2 u^4 = 0, \text{ or } \nabla \cdot \vec{B} = 0 \quad (44)$$

is used to identify the steady effects from the dynamic process.

Based on this strategy, when the current density is introduced, the matter must be deformable. For the general cases, there is an additional motion equation [1]:

$$\frac{\lambda + \mu}{c^2}\frac{\partial^2 u^4}{(\partial t)^2} + \mu[\frac{\partial^2 u^4}{(\partial x^1)^2} + \frac{\partial^2 u^4}{(\partial x^2)^2} + \frac{\partial^2 u^4}{(\partial x^3)^2}] = 0 \quad (45)$$

It can be rewritten as:

$$\frac{\lambda + \mu}{c^2}\frac{\partial A}{\partial t} + \mu \nabla \cdot \vec{B} = 0 \quad (46)$$

Note that if one defines $(\lambda + \mu)A$ as scalar potential for electric field, $\mu\vec{B}$ as the vector potential for magnetic field, the equation (46) is the Lorentz gauge. This equation can be established from the contra-covariant form of matter motion equation derived from the least action principle for the non-commutative field.

Therefore, Maxwell equations and both Lorentz gauge and Coulomb gauge [10] are derived in this research.

The above equations form the Maxwell electromagnetic field equations.

*(b). Coherent Light*

If $u^4$ is harmonic about time $t$ with frequency $\omega$ and wave number $k_i$, one will has:

$$A = \frac{\partial u^4}{\partial t} = \omega u_0^4, \quad B_i = \frac{\partial u^4}{\partial x^i} = k_i u_0^4 \tag{47}$$

Putting them into equation (37), one finds that, for $i=1,2,3$:

$$(\lambda + \mu)(\frac{\partial \omega}{\partial x^i} + \omega k_i) = -\mu(\frac{\partial k_i}{\partial t} + \omega k_i) \tag{48}$$

That is:

$$(\lambda + \mu)\frac{\partial \omega}{\partial x^i} + \mu \frac{\partial k_i}{\partial t} = -(\lambda + 2\mu)\omega k_i \tag{49}$$

The intrinsic feature of wave number-frequency relation is that it is completely determined by the charge and mass ratio. In physics, for basic particles, the charge-mass indeed plays as basic parameter. This is clear from following discussion.

If $\omega$ is a spatial constant, then one has:

$$\frac{\partial k_i}{\partial t} = -\frac{\lambda + 2\mu}{\mu}\omega k_i = -\frac{\tilde{\rho}}{\rho}\omega k_i \tag{50}$$

It has a solution:

$$k_i = k_i^0 \exp(-\frac{\tilde{\rho}}{\rho}\omega t) \tag{51}$$

It shows that the wave number decreases to zero for long time duration. It shows that the matter field is localized about time. It can be used to define the life-duration of basic particles.

If $k_i$ is a time constant, then one has:

$$\tilde{q}\frac{\partial \omega}{\partial x^i} = -\tilde{\rho}\omega k_i \tag{52}$$

It has a solution:

$$\omega = \omega_0 \exp(-\frac{\tilde{\rho}}{\tilde{q}}k_i x^i) \tag{53}$$

It shows that the matter field is localized about space, so for such a kind of matter the limit configuration can be defined.

For the light wave in vacuum, as the physical observation has proven that the wave number-frequency is linear relation, based on the above equations, there must be equation:

$$\lambda + 2\mu = 0, \text{ or } \tilde{q} + \rho = 0 \tag{54}$$

It means that the light is a special matter in that the addition of its Newtonian mass and its Coulomb electric charge is zero. In fact, this conclusion is true for the electromagnetic wave in vacuum.

The equations (48-53) in fact can be used to explain the theoretic bases of coherent light or the coherent wave process of electromagnetic wave [11].

*(c). Pure Magnetic Field (and Geo-Magnetic Field)*

If $\frac{\partial A}{\partial x^i} = 0$ or $\lambda + \mu = 0$, that is if the $A$ is a pure time function, one will get:

$$f^1 = \mu \frac{\partial^2 u^4}{\partial x^1 \partial t} = \rho c^2 \frac{\partial B_1}{\partial t}$$

$$f^2 = \mu \frac{\partial^2 u^4}{\partial x^2 \partial t} = \rho c^2 \frac{\partial B_2}{\partial t} \qquad (55)$$

$$f^3 = \mu \frac{\partial^2 u^4}{\partial x^3 \partial t} = \rho c^2 \frac{\partial B_3}{\partial t}$$

It defines magnetic field. So, one can say that the time gradient of the spatial distribution of time displacement determines magnetic field. When the spatial force has vorticity, then it will be the source of magnetic variation. That is:

$$\nabla \times \vec{f} = \rho c^2 \frac{\partial \vec{H}}{\partial t} \qquad (56)$$

It shows that the vorticity of Newtonian spatial force is proportional with the magnetic field time derivative and the coefficient is the Newtonian mass.

This equation can be used to explain the Earth magnetic field variation and the source of the Geo-magnetic field. The existence of magnetic field of stars should be normal phenomena.

This equation can be checked by the related observation in space physics, especially the Sun magnetic field..

*(d). Pure Electric Field*

If $\frac{\partial B_i}{\partial t} = 0$ for $i = 1,2,3$, that is if the $B_i$ is a pure spatial vector-function, one will get:

$$f^1 = (\lambda + \mu) \frac{\partial^2 u^4}{\partial t \partial x^1} = \tilde{q} c^2 \frac{\partial A}{\partial x^1}$$

$$f^2 = (\lambda + \mu) \frac{\partial^2 u^4}{\partial t \partial x^2} = \tilde{q} c^2 \frac{\partial A}{\partial x^2} \qquad (57)$$

$$f^3 = (\lambda + \mu) \frac{\partial^2 u^4}{\partial t \partial x^3} = \tilde{q} c^2 \frac{\partial A}{\partial x^3}$$

$$f^4 = \frac{\lambda + \mu}{c} \frac{\partial^2 u^4}{(\partial t)^2} = \tilde{q} c \frac{\partial A}{\partial t}$$

This field corresponds to a conservative field in four-dimensional coordinator system.

Without outside source, the variation of cosmic force $f^4$ will produce an additional electric potential. This additional electric potential will produce spatial forces.

If one defines this as weak action force, the field described by equation (20) may be defined as strong action.

Summing up above discussion about time displacement field, one finds that gravity field, electro-magnetic field, and quantum field are time gradient field. They are related with Newton's mechanics in intrinsic sense. So, the united field theory of matter motion can be expressed by the finite geometrical field theory given by this paper.

**5. Geometrical Invariants and Conservative Fields in Classical Physics**

For a matter, if it is defined initially in vacuum geometry as having geometric invariant:

$$ds_0^2 = g_{11}^0 (dx^1)^2 + g_{22}^0 (dx^2)^2 + g_{33}^0 (dx^3)^2 + c^2 (dx^4)^2 \qquad (58)$$

Then, the matter motion in purely time displacement sense will have current geometric invariant as:

$$ds^2 = [g_{11}^0 + (c\frac{\partial u^4}{\partial x^1})^2](dx^1)^2 + [g_{22}^0 + (c\frac{\partial u^4}{\partial x^2})^2](dx^2)^2 + [g_{33}^0 + (c\frac{\partial u^4}{\partial x^3})^2](dx^3)^2$$
$$+ 2c^2 \frac{\partial u^4}{\partial x^1}\frac{\partial u^4}{\partial x^2} dx^1 dx^2 + 2c^2 \frac{\partial u^4}{\partial x^2}\frac{\partial u^4}{\partial x^3} dx^2 dx^3 + 2c^2 \frac{\partial u^4}{\partial x^3}\frac{\partial u^4}{\partial x^1} dx^3 dx^1$$
$$+ [V^2 + c^2 + 2c^2 \frac{\partial u^4}{\partial x^4} + c^2(\frac{\partial u^4}{\partial x^4})^2](dx^4)^2 \tag{59}$$

Here, $V$ is the velocity of inertial system. The physical geometric invariant is:

$$ds^2 - ds_0^2 = (c\frac{\partial u^4}{\partial x^1})^2(dx^1)^2 + (c\frac{\partial u^4}{\partial x^2})^2(dx^2)^2 + (c\frac{\partial u^4}{\partial x^3})^2(dx^3)^2$$
$$+ 2c^2 \frac{\partial u^4}{\partial x^1}\frac{\partial u^4}{\partial x^2} dx^1 dx^2 + 2c^2 \frac{\partial u^4}{\partial x^2}\frac{\partial u^4}{\partial x^3} dx^2 dx^3 + 2c^2 \frac{\partial u^4}{\partial x^3}\frac{\partial u^4}{\partial x^1} dx^3 dx^1$$
$$+ [V^2 + 2c^2 \frac{\partial u^4}{\partial x^4} + c^2(\frac{\partial u^4}{\partial x^4})^2](dx^4)^2 \tag{60}$$

So, geometric invariant for physical motion can be defined as:

$$d\Sigma^2 = ds^2 - ds_0^2 = c^2(du^4)^2 + 2c^2 \frac{\partial u^4}{\partial x^4}(dx^4)^2 + V^2(dx^4)^2 \tag{61}$$

where:

$$(du^4)^2 = (\frac{\partial u^4}{\partial x^1})^2(dx^1)^2 + (\frac{\partial u^4}{\partial x^2})^2(dx^2)^2 + (\frac{\partial u^4}{\partial x^3})^2(dx^3)^2 + (\frac{\partial u^4}{\partial x^4})^2(dx^4)^2$$
$$+ 2\frac{\partial u^4}{\partial x^1}\frac{\partial u^4}{\partial x^2} dx^1 dx^2 + 2\frac{\partial u^4}{\partial x^2}\frac{\partial u^4}{\partial x^3} dx^2 dx^3 + 2\frac{\partial u^4}{\partial x^3}\frac{\partial u^4}{\partial x^1} dx^3 dx^1 \tag{62}$$

From physical consideration, no matter what initial geometry system is used as initial co-moving coordinator system geometry, the subjectivity of physical motion will require the $d\Sigma^2$ be geometrical invariant. It can be divided into three simple typical cases, as described bellow.

*(a). Non-Quantum Matter Motion (Newton Mass-point)*

Based on the previous research, for non-quantum matter motion the time displacement field component can be replaced by Newton's Acceleration Law. It is equivalent to suppose there is no time displacement. Hence, we suppose for non-quantum matter motion:

$$V^2 >> 2c^2 \frac{\partial u^4}{\partial x^4} + c^2(du^4)^2 \tag{63}$$

Hence, the geometrical invariant of non-quantum matter motion is:

$$d\Sigma^2 = V^2(dx^4)^2 \tag{64}$$

That is:

$$d\Sigma^2 = (du^1)^2 + (du^2)^2 + (du^3)^2 \tag{65}$$

This defines the conventional three-dimensional physical measure space. It defines the spatial motion distance as geometrical invariant. For Newton's matter point, it leads to the kinetic energy conservation. By equation (64), for mass $\rho$ distribution, the spatial integral:

$$W = \int \rho d\Sigma^2 d\Omega = \int \rho V^2 (dx^4)^2 d\Omega \tag{66}$$

is geometrical invariant. It leads to the kinetic energy conservation:

$$w = \int \rho V^2 d\Omega \tag{67}$$

This is the traditional Newtonian mechanics. The space is Cartesian and the time is an independent parameter.

*(b). Very High Speed Motion of Newtonian Matter*

For very high speed matter motion, the time gradient of time displacement can be ignored, that is to say:

$$V^2 \gg 2c^2 \frac{\partial u^4}{\partial x^4} + c^2 (\frac{\partial u^4}{\partial x^4})^2 \tag{68}$$

Then we will have the geometrical invariant as:

$$d\Sigma^2 = ds^2 - ds_0^2 = c^2 (d\tilde{u}^4)^2 + (du^1)^2 + (du^2)^2 + (du^3)^2 \tag{69}$$

where:

$$(d\tilde{u}^4)^2 = (\frac{\partial u^4}{\partial x^1})^2 (dx^1)^2 + (\frac{\partial u^4}{\partial x^2})^2 (dx^2)^2 + (\frac{\partial u^4}{\partial x^3})^2 (dx^3)^2$$
$$+ 2\frac{\partial u^4}{\partial x^1}\frac{\partial u^4}{\partial x^2} dx^1 dx^2 + 2\frac{\partial u^4}{\partial x^2}\frac{\partial u^4}{\partial x^3} dx^2 dx^3 + 2\frac{\partial u^4}{\partial x^3}\frac{\partial u^4}{\partial x^1} dx^3 dx^1 \tag{70}$$

It can be seen that $d\tilde{u}^4$ is independent with $dx^4$, so it is a scalar defined in three-dimensional space.

For matter behaves as the time displacement is defined by the following wave form:

$$u^4 = U_0 \exp[(\vec{R} \cdot \vec{k} \pm \frac{c}{n} t) j] \tag{71}$$

or matter behaves as the time displacement is defined by the following spatial-localized harmonic particle form:

$$u^4 = U_0 \exp(\vec{R} \cdot \vec{l} \pm \frac{c}{n} t j) \tag{72}$$

If we take the definition of:

$$d\tilde{u}^4 = j d\tau \tag{73}$$

and explain the $d\tau$ as the universal time (world time), we will get the geometrical invariant as:

$$d\Sigma^2 = c^2 (d\tilde{u}^4)^2 + (du^1)^2 + (du^2)^2 + (du^3)^2 \tag{74}$$

That is:

$$d\Sigma^2 = (du^1)^2 + (du^2)^2 + (du^3)^2 - c^2 (d\tau)^2 \tag{75}$$

In general relativity theory and in gauge field theory, this equation form is used to define the geometry of space-time continuum under the meaning that the $d\Sigma^2$ is the geometrical invariant of "word-line distance". However, based on this paper's research, this is only true when the time displacement is normal and in form (71) or (72). For more complicated matter motion, the gauge field theory and general relativity will be failed.

Summing up above discussion, we can see that traditional physical conservation laws and the geometrical invariant introduced by the gauge field theory and general relativity are included in the finite geometrical field theory as simple special cases.

*(c). Conservative Fields*

Based on the previous research, for quantum matter motion the time displacement field component play the main roles. It is equivalent to suppose there is a significant time displacement. Hence, we suppose for quantum matter motion:

$$V^2 + 2c^2 \frac{\partial u^4}{\partial x^4} << c^2 (du^4)^2 \qquad (68)$$

Hence, the geometrical invariant of non-quantum matter motion is:

$$d\Sigma^2 = ds^2 - ds_0^2 = c^2 (du^4)^2 \qquad (69)$$

So, we can define time displacement is a scalar function in four-dimensional time-space. As a special case, this geometrical invariant can be explained as matter-energy conservation.

In the paper "Inertial System and Special Relativity"[1], the conservative field (gravitation field and static electric field) has a typical solution in form:

$$\frac{\partial u^4}{\partial x^4} = \frac{M}{r} \qquad (70)$$

where, the zero point is usually defined by the center point $(x_0^1, x_0^2, x_0^3)$ of the matter; $M$ parameter may be related to Newtonian mass of the matter or the static charge of the matter, or their summation, which depends on the topic. Here, it is simply refereed as general charge. In this case, the time displacement is the only one non-zero displacement component.

According to the motion transformation in is case (referring equation (1)), one has:

$$g_{11} = 1 + (u^4|_1)^2 \cdot c^2$$

$$g_{22} = 1 + (u^4|_2)^2 \cdot c^2$$

$$g_{33} = 1 + (u^4|_3)^2 \cdot c^2 \qquad (71)$$

$$g_{44} = V^2 + (1 + u^4|_4)^2 \cdot c^2$$

For simplicity, $g_{ij}, i \neq j$ are not taken into consideration. (This topic will be fully treated in the part three paper of this research, where quantum mechanics is the main theme).

For an inertial system defined on the conservative matter and the constant general charge, $V = 0$, putting the general solution (70) into the equation (71), one has:

$$g_{11} = 1 + (\frac{x^1 - x_0^1}{r^2})^2 \cdot M^2 (x^4 - x_0^4)^2 c^2$$

$$g_{22} = 1 + (\frac{x^2 - x_0^2}{r^2})^2 \cdot M^2 (x^4 - x_0^4)^2 c^2$$

$$g_{33} = 1 + (\frac{x^3 - x_0^3}{r^2})^2 \cdot M^2 (x^4 - x_0^4)^2 c^2 \qquad (72)$$

$$g_{44} = (1 + \frac{M}{r})^2 \cdot c^2 = (1 + \frac{2M}{r} + \frac{M^2}{r^2}) \cdot c^2$$

In fact omitting the higher order infinitesimals, the approximation is:

$$g_{11} = g_{22} = g_{33} = 1,$$

$$g_{44} = (1 + \frac{2M}{r}) \cdot c^2 \qquad (73)$$

For gravity field, the equation (73) is approximation of the typical solution Schwarzschild metric. The equation (72) is the approximation of Kerr metric.

To make this point clear, let's consider the requirement of inertial system. For this case, the physical time scale is defined by: $g_{44} = g_{44}^0 = c^2$, to make the space scale be the absolute same (not dependent on $(x^4 - x_0^4)$), the space scale must be adjusted to make: $g_{(ii)} g_{44}$ is the absolute same. So, the current gauge that meets the definition of inertial system and the conservative field equations can be defined as:

$$\tilde{g}_{11} = \tilde{g}_{22} = \tilde{g}_{33} = (1 + \frac{2M}{r})^{-1},$$
$$\tilde{g}_{44} = (1 + \frac{2M}{r}) \cdot c^2$$
(74)

However, for the cosmic ages old events, the spatial curvature may be cannot be ignored when the following the $(x^4 - x_0^4)^2$ is very big, that is:

$$g_{(ii)} = 1 + (\frac{x^i - x_0^i}{r^2})^2 \cdot M^2 (x^4 - x_0^4)^2 c^2 \approx 1 + \vartheta$$
(75)

On this sense, the oldest gravity field has the largest curvature of space. This point is very intrinsic for astronomy matters. Therefore, it is easy to understand that the general relativity theory is well accepted by astronomy physics.

Unfortunately, the effects caused by the cosmic age dependent curvature of space are not taken into consideration for the observation cosmics or astronomy physics. This may be the main cause of introducing the Big-ban theory and its related alternatives.

For the micro-world matter, when the electric charge is concerned, the old matter still has a much curvature space. This makes the quantum mechanics become the most complicated physical world, where the general relativity still play an important role. This will be the main topic of next paper on finite geometrical field theory.

## 6. Conclusion

When the matter has no macro spatial motion, such as gravity field or electronic field, the time displacement field is introduced to describe such a kind of matter motion. For commutative case, it is founded that gravity field, electro-magnetic field, and quantum field are time gradient field. The absolute electrical charge quantity may be expressed as $\tilde{q} = (\lambda + \mu)/c^2$. When there is no electrical charge quantity, the gravity mass and the inertial mass in Newton's mechanics are the same. The matter has three typical existing forms: traveling wave, localized harmonic vibrating, and exponential expanding or decaying. The commutable matter motion defines conservative field. The commutative matter motion defines a quantum field charge $\tilde{\rho} = (\lambda + 2\mu)/c^2$ (defined as matter complete quality).

For non-commutative matter motion, the Maxwell electromagnetic field equations are established, based on the covariant form of matter motion equation derived from the least action principle. When the contra-covariant is required for the non-commutative field, both Lorentz gauge and Coulomb gauge are derived in this research. The paper shows that the light is a special matter in that the addition of its Newtonian mass and its Coulomb electric charge is zero. In fact, this conclusion is true for the electromagnetic wave in vacuum

For the conservative field, the research shows that once the mass density and the Coulom charge dendity are given, the macro spacetime feature is completely determined.Both of them are intrisinc features of macro matter in cosimic background.

However, for the cosmic ages old events, the spatial curvature may be cannot be ignored when the following the $(x^4 - x_0^4)^2$ is very big, that is:

$$g_{(ii)} = 1 + (\frac{x^i - x_0^i}{r^2})^2 \cdot M^2 (x^4 - x_0^4)^2 c^2 \approx 1 + \vartheta \qquad (75)$$

On this sense, the oldest gravity field has the largest curvature of space. This point is very intrinsic for astronomy matters.